\documentclass[twocolumn,aps,nofootinbib,prl,groupedaddress]{revtex4}
\usepackage{epsfig,amssymb,amsmath}
 \def\labell#1{\label{#1}}
\def\section#1{{\em #1:--- }} 
\def\togli#1{}
\def\>{\rangle}
\def\<{\langle}

\usepackage{color}

\begin{document}

\title{ Quantum measurement bounds beyond the uncertainty relations}

\author{Vittorio Giovannetti$^1$, Seth Lloyd$^2$, Lorenzo Maccone$^3$}
\affiliation{ $^1$ NEST, Scuola Normale Superiore and Istituto
  Nanoscienze-CNR,
  piazza dei Cavalieri 7, I-56126 Pisa, Italy \\
  $^2$Dept.~of Mechanical Engineering, Massachusetts Institute of
  Technology, Cambridge, MA 02139, USA  \\
  $^3$Dip.~Fisica ``A.~Volta'', INFN Sez.~Pavia, Universit\`a di
  Pavia, via Bassi 6, I-27100 Pavia, Italy}

\begin{abstract}
  We give a bound to the precision in the estimation of a parameter in
  terms of the expectation value of an observable. It is an extension
  of the Cram\'er-Rao inequality and of the Heisenberg uncertainty
  relation, where the estimation precision is typically bounded in
  terms of the variance of an observable.
\end{abstract}
\maketitle 

Quantum measurements are limited by bounds such as the Heisenberg
uncertainty relations \cite{heisenberg,robertson} or the quantum
Cram\'er-Rao inequality \cite{holevo,helstrom,BRAU96,BRAU94}, which
typically constrain the ability in recovering a target quantity
(e.g.~a relative phase) through the {\em standard deviation} of a
conjugate one (e.g.~the energy) evaluated on the state of the probing
system.  Here we give a new bound related to the {\em expectation
  value}: we show that the precision in the quantity cannot scale
better than the inverse of the expectation value (above a ``ground
state'') of its conjugate counterpart. It is especially relevant in
the expanding field of quantum metrology \cite{review}: it settles in
the positive the longstanding conjecture of quantum optics
\cite{caves,yurke,barry,ou,bollinger,smerzi}, recently challenged
\cite{dowling,rivasluis,zhang}, that the ultimate phase-precision
limit in interferometry is lower bounded by the inverse of the total
number of photons employed in the estimation process.
  
The aim of Quantum Parameter
Estimation~\cite{holevo,helstrom,BRAU96,BRAU94} is to recover the {\em
  unknown} value $x$ of a parameter that is written into the state
$\rho_x$ of a probe system through some {\em known} encoding mechanism
$U_x$. For example, we can recover the relative optical delay $x$
among the two arms of a Mach-Zehnder interferometer described by its
unitary evolution $U_x$ using as probe a light beam fed into the
interferometer. The statistical nature of quantum mechanics induces
fluctuations that limit the ultimate precision which can be achieved
(although we can exploit quantum ``tricks'' such as entanglement and
squeezing in optimizing the state preparation of the probe and/or the
detection stage \cite{GIOV06}).
In particular, if the encoding stage is repeated
several times using $\nu$ identical copies of the same probe input
state $\rho_x$, the root mean square error (RMSE) $\Delta X$ of the
resulting estimation process is limited by the quantum Cram\'er-Rao
bound~\cite{holevo,helstrom,BRAU96,BRAU94} $\Delta X\geqslant
1/\sqrt{\nu Q(x)}$, where ${Q}(x)$ is the quantum Fisher information.
For pure probe states and unitary encoding mechanism $U_x$, ${Q}(x)$
is equal to the variance $(\Delta H)^2$ (calculated on the probe
state) of the generator $H$ of the transformation $U_x=e^{-ixH}$.
In this case, the Cram\'er-Rao bound takes the form 
\begin{eqnarray}
  \Delta
  X\geqslant 1/(\sqrt{\nu} \Delta H)\label{QC}\;
\end{eqnarray}
of an uncertainty relation \cite{BRAU94,BRAU96}. In
fact, if the parameter $x$ can be connected to an observable,
Eq.~\eqref{QC} corresponds to the Heisenberg uncertainty relation for
conjugate variables~\cite{heisenberg,robertson}. This bound is
asymptotically achievable in the limit of $\nu \rightarrow \infty$
\cite{holevo,helstrom}.

\begin{figure}[t]
\begin{center}
\epsfxsize=.6\hsize\leavevmode
\epsffile{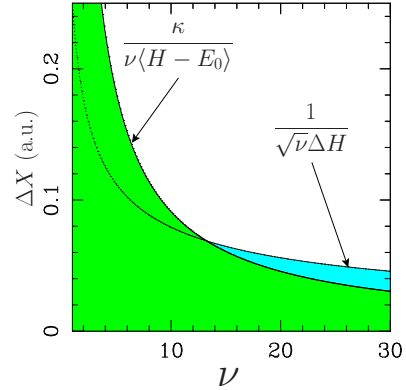}
\end{center}
\caption{ Lower bounds to the precision estimation $\Delta X$ as a
  function of the experimental repetitions $\nu$. The green area in
  the graph represents the forbidden values due to our bound
  \eqref{ris}. The blue (dashed-line) area represents the forbidden
  values due to the Cram\'er-Rao bound, or the Heisenberg uncertainty,
  \eqref{QC}. Possible estimation strategies have precision $\Delta
  X$ that cannot penetrate in the colored regions. For large $\nu$ the
  Cram\'er-Rao bound (which scales as $1/\sqrt{\nu}$) is stronger, as
  expected since in this regime it is achievable. Our bound is not
  achievable in general, so that the green area may be expanded when
  considering specific estimation strategies.  [Here we used $\langle
  H\rangle-E_0=0.1$ (a.u.) and $\Delta H=4$ (a.u.).]}
\labell{f:compar}
\end{figure}

Here we will derive a bound in terms of the expectation value of $H$,
which (in the simple case of constant $\Delta X$) takes the form (see
Fig.~\ref{f:compar})
\begin{eqnarray}
\Delta X\geqslant \kappa/[\nu(\langle H\rangle-E_0)]
\labell{ris}\;,
\end{eqnarray}
where $E_0$ is the value of a ``ground state'', the minimum eigenvalue
of $H$ whose eigenvector is populated in the probe state (e.g.~the
ground state energy when $H$ is the probe's Hamiltonian), and $\kappa\simeq
0.091$ is a constant of order one. Our bound holds both for biased and
unbiased measurement procedures, and for pure and mixed probe states.
When $\Delta X$ is dependent on $x$, a constraint of the
form~(\ref{ris}) can be placed on the average value of $\Delta X(x)$
evaluated on any two values $x$ and $x'$ of the parameter which are
sufficiently separated, namely
\begin{eqnarray} \label{bd1} \frac{\Delta X(x) + \Delta X(x')}{2} &\geqslant
  & \frac{\kappa}{\nu(\<H\> - E_0)} \;.
\end{eqnarray} 
Hence, we cannot exclude that strategies whose error $\Delta X$ depend
on $x$ may have a ``sweet spot'' where the bound \eqref{ris} may be
beaten \cite{rivasluis}, but inequality \eqref{bd1} shows that the
average value of $\Delta X$ is subject to the bound.  Thus, these
strategies are of no practical use, since the sweet spot depends on
the unknown parameter $x$ to be estimated and the extremely good
precision in the sweet spot must be counterbalanced by a
correspondingly bad precision nearby.

Proving the bound~\eqref{ris} in full generality is clearly not a
trivial task since no definite relation can be established between
$\nu(\langle H\rangle-E_0)$ and the term $\sqrt{\nu} \Delta H$ on
which the Cram\'er-Rao bound is based. In particular, scaling
arguments on $\nu$ cannot be used since, on one hand, the value of
$\nu$ for which Eq.~(\ref{QC}) saturates is not known (except in the
case in which the estimation strategy is fixed \cite{caves}, which has
little fundamental relevance) and, on the other hand, input probe
states $\rho$ whose expectation values $\langle H \rangle$ depend
explicitly on $\nu$ may be employed, e.g.~see Ref.~\cite{rivasluis}.
To circumvent these problems our proof is based on the quantum speed
limit~\cite{qspeed}, a generalization of the Margolus-Levitin
~\cite{margolus} and Bhattacharyya bounds~\cite{bhatta,man} which
links the fidelity $F$ between the two joint states
$\rho_x^{\otimes\nu}$ and $\rho_{x'}^{\otimes\nu}$ to the difference
$x'-x$ of the parameters $x$ and $x'$ imprinted on the states through
the mapping $U_x=e^{-ixH}$ [The fidelity between two states $\rho$ and
$\sigma$ is defined as $F=\{\mbox{Tr}[\sqrt{\sqrt{\rho}
  \sigma\sqrt{\rho}}]\}^2$. A connection between quantum metrology and
the Margolus-Levitin theorem was proposed in \cite{kok}, but this
claim was subsequently retracted in \cite{erratum}.]  In the case of
interest here, the quantum speed limit \cite{qspeed} implies
\begin{eqnarray} 
  |x'-x|\geqslant\frac\pi2
  \max  \left[\frac{\alpha(F)}{\nu(\<H\>-E_0)}\;,\
    \frac{\beta(F)}{\sqrt{\nu} \Delta H}\right]  \;\labell{newqsl}\;,
\end{eqnarray}
where the $\nu$ and $\sqrt{\nu}$ factors at the denominators arise
from the fact that here we are considering $\nu$ copies of the probe
states $\rho_x$ and $\rho_{x'}$, and where
$\alpha(F)\simeq\beta^2(F)=4\arccos^2(\sqrt{F})/\pi^2$ are the
functions plotted in Fig.~\ref{f:qsl} of the supplementary material.
The inequality~\eqref{newqsl} tells us that the parameter difference
$|x'-x|$ induced by a transformation $e^{-i(x'-x)H}$ which employs
resources $\<H\>-E_0$ and $\Delta H$ cannot be arbitrarily small (when
the parameter $x$ coincides with the evolution time, this sets a limit
to the ``speed'' of the evolution, the quantum speed limit).

We now give the main ideas of the proof of \eqref{ris} by focusing on
a simplified scenario, assuming pure probe states $|\psi_x\rangle=U_x
|\psi\rangle$, and unbiased estimation strategies constructed in terms
of projective measurements with RSME $\Delta X$ that do not depend on
$x$ (all these assumptions are dropped in the supplementary material).
For unbiased estimation, $x=\sum_j P_j(x) x_j$ and the RMSE coincides
with the variance of the distribution $P_j(x)$, i.e.~$\Delta
X=\sqrt{\sum_j P_j(x) [ x_j-x]^2}$, where $P_j(x) = |\langle x_j |
\psi_x \rangle^{\otimes\nu}|^2$ is the probability of obtaining the
result $x_j$ while measuring the joint state
$|\psi_x\rangle^{\otimes\nu}$ with a projective measurement on the
joint basis $|x_j\rangle$. Let us consider two values $x$ and $x'$ of
the parameter that are further apart than the measurement's RMSE,
i.e.~$x'-x=2 \lambda \Delta X$ with $\lambda>1$.  If no such $x$ and
$x'$ exist, the estimation is extremely poor: basically the whole
domain of the parameter is smaller than the RMSE.  Hence, for
estimation strategies that are sufficiently accurate to be of
interest, we can always assume that such a choice is possible (see
below).  The Tchebychev inequality states that for an arbitrary
probability distribution $p$, the probability that a result $x$ lies
more than $\lambda\Delta X$ away from the average $\mu$ is upper
bounded by $1/\lambda^2$, namely $p(|x-\mu|\geqslant \lambda\Delta
X)\leqslant 1/\lambda^2$. It implies that the probability that
measuring $|\Psi_{x'}\rangle :=|\psi_{x'}\rangle^{\otimes\nu}$ the
outcome $x_j$ lies within $\lambda\Delta X$ of the mean value
associated with $|\Psi_x\rangle:=|\psi_x\rangle^{\otimes\nu}$ cannot
be larger $1/\lambda^2$. By the same reasoning, the probability that
measuring $|\Psi_{x}\rangle$ the outcome $x_j$ will lie within
$\lambda\Delta X$ of the mean value associated with
$|\Psi_{x'}\rangle$ cannot be larger $1/\lambda^2$.  This implies that
the overlap between the states $|\Psi_{x}\rangle$ and
$|\Psi_{x'}\rangle$ cannot be too large: more precisely, $F=|
\langle\Psi_x|\Psi_{x'} \rangle|^2\leqslant 4/\lambda^2$. Replacing
this expression into \eqref{newqsl} (exploiting the fact that $\alpha$
and $\beta$ are decreasing functions) we obtain
\begin{eqnarray}
2\lambda  \Delta X\geqslant
  \frac\pi{2}
   \max  \left[\frac{\alpha(4/\lambda^2)}{\nu(\<H\>-E_0)}\;,\
    \frac{\beta(4/\lambda^2)}{\sqrt{\nu} \Delta H}\right]
  \labell{ineq}\;,
\end{eqnarray}
whence we obtain \eqref{ris} by optimizing over $\lambda$ the first
term of the $\max$, i.e.~choosing
$\kappa=\sup_\lambda\pi\:\alpha(4/\lambda^2)/(4\lambda)\simeq 0.091$. The
second term of the $\max$ gives rise to a quantum Cram\'er-Rao type
uncertainty relation (or a Heisenberg uncertainty relation) which,
consistently with the optimality of Eq.~(\ref{QC}) for $\nu\gg1$, has
a pre-factor $\pi \beta(4/\lambda^2)/ (4 \lambda)$ which is smaller
than $1$ for all $\lambda$. This means that for large $\nu$ the bound
\eqref{ris} will be asymptotically superseded by the Cram\'er-Rao
part, which scales as $\propto 1/\sqrt{\nu}$ and is achievable in this
regime.

Analogous results can be obtained (see supplementary material) when
considering more general scenarios where the input states of the
probes are not pure, the estimation process is biased, and it is
performed with arbitrary POVM measurements. (In the case of biased
measurements, the constant $\kappa$ in \eqref{ris} and \eqref{bd1}
must be replaced by $\kappa= \sup_{\lambda} \pi
\alpha(4/\lambda^2)/[4(\lambda+1)]\simeq 0.074$, where a $+1$ term
appears in the denominator.) In this generalized context, whenever the
RMSE depends explicitly on the value $x$ of the parameter, the
result~\eqref{ris} derived above is replaced by the weaker
relation~\eqref{bd1}. Such inequality clearly does not necessarily
exclude the possibility that at a ``sweet spot'' the estimation might
violate the scaling~(\ref{ris}).  However, Eq.~(\ref{bd1}) is still
sufficient strong to exclude accuracies of the form $\Delta X(x)
=1/R(x,\nu\langle H\rangle)$ where, as in Refs.~\cite{ssw,rivasluis},
$R(x,z)$ is a function of $z$ which, for all $x$, increases more than
linearly, i.e.~$\lim_{z\rightarrow \infty} z/R(x,z)=0$.

The bound~\eqref{ris} has been derived under the explicit assumption
that $x$ and $x'$ exists such that $x'-x\geqslant 2 \lambda \Delta X$
for some $\lambda >1$, which requires one to have $x'-x\geqslant 2
\Delta X$.  This means that the estimation strategy must be good
enough: the probe is sufficiently sensitive to the transformation
$U_x$ that it is shifted by more than $\Delta X$ during the
interaction. The existence of pathological estimation strategies which
violate such condition cannot be excluded {\em a priori}. Indeed
trivial examples of this sort can be easily constructed, a fact which
may explain the complicated history of the Heisenberg bound with
claims \cite{caves,yurke,barry,ou,bollinger,smerzi} and counterclaims
\cite{dowling,rivasluis,zhang,ssw}. It should be stressed however,
that the assumption $x'-x\geqslant 2 \Delta X$ is always satisfied
except for extremely poor estimation strategies with such large errors
as to be practically useless. One may think of repeating such a poor
estimation strategy $\nu>1$ times and of performing a statistical
average to decrease its error.  However, for sufficiently large $\nu$
the error will decrease to the point in which the $\nu$ repetitions of
the poor strategy are, collectively, a good strategy, and hence again
subject to our bounds \eqref{ris} and \eqref{bd1}.

Our findings are particularly relevant in the field of quantum optics,
where a controversial and longly debated problem
\cite{caves,yurke,barry,ou,bollinger,smerzi,ssw,dowling,rivasluis,zhang}
is to determine the scaling of the ultimate limit in the
interferometric precision of estimating a phase as a function of the
energy $\langle H\rangle$ devoted to preparing the $\nu$ copies of the
probes: it has been conjectured
\cite{caves,yurke,barry,ou,bollinger,smerzi} that the phase RMSE is
lower bounded by the inverse of the total number of photons employed
in the experiment, the ``Heisenberg bound'' for
interferometry\footnote{This ``Heisenberg''
  bound~\cite{caves,yurke,barry,ou,bollinger,smerzi} should not be
  confused with the Heisenberg scaling defined for general quantum
  estimation problem~\cite{review} in which the $\sqrt{\nu}$ at the
  denominator of Eq.~(\ref{QC}) is replaced by $\nu$ by feeding the
  $\nu$ inputs with entangled input states -- e.g. see
  Ref.~\cite{review,GIOV06}.}. Its achievability has been recently
proved \cite{HAYA10-1}, and, in the context of quantum parameter
estimation, it corresponds to an equation of the form of
Eq.~\eqref{ris}, choosing $x=\phi$ (the relative phase between the
modes in the interferometer) and $H=a^\dag a$ (the number operator).
The validity of this bound has been questioned several times
\cite{ssw,dowling,rivasluis,zhang}. In particular schemes have been
proposed~\cite{ssw,rivasluis} that apparently permit better scalings
in the achievable RMSE (for instance $\Delta X \approx (\nu\langle
H\rangle)^{-\gamma}$ with $\gamma>1$).  None of these protocols have
conclusively proved such scalings for arbitrary values of the
parameter $x$, but a sound, clear argument against the possibility of
breaking the $\gamma=1$ scaling of Eq.~(\ref{ris}) was missing up to
now. Our results validate the Heisenberg bound by showing that it
applies to all those estimation strategies whose RMSE $\Delta X$ do
not depend on the value of the parameter $x$, and that the remaining
strategies can only have good precision for isolated (hence
practically useless) values of the unknown parameter $x$.\newline

V.G. acknowledges support by MIUR through FIRB-IDEAS Project No.
RBID08B3FM. S.L. acknowledges Intel, Lockheed Martin, DARPA, ENI under
the MIT Energy Initiative, NSF, Keck under xQIT, the MURI QUISM
program, and Jeffrey Epstein.  L.M.  acknowledges useful discussions
with Prof. Alfredo Luis.

\subsection*{Supplementary material}

Our bound refers to the estimation of the value $x$ of a real
parameter $X$ that identifies a unitary transformation $U_x=e^{-iHx}$,
generated by an Hermitian operator $H$.  The usual setting in quantum
channel parameter estimation~(see \cite{review} for a recent review)
is to prepare $\nu$ copies of a probe system in a fiducial state
$\rho$, apply the mapping $U_x$ to each of them as
$\rho\to\rho_x=U_x\rho {U_x}^\dag$, and then perform a (possibly
joint) measurement on the joint output state $\rho_x^{\otimes \nu}$,
the measurement being described by a generic Positive Operator-Valued
Measure (POVM) of elements $\{ E_j\}$. [The possibility of applying a
joint transformation on the $\nu$ probes before the interaction $U_x$
(e.g.~to entangle them as studied in \cite{GIOV06}) can also be
considered, but it is useless in this context, since it will not
increase the linear scaling in $\nu$ of the term $\nu(\langle
H\rangle-E_0)$ that governs our bounds.] The result $j$ of the measurement is finally
used to recover the quantity $x$ through some data processing which
assigns to each outcome $j$ of the POVM a value $x_j$ which represents
the estimation of $x$.  The accuracy of the process can be gauged by
the RMSE of the problem, i.e.~by the quantity
\begin{eqnarray}\label{defdelta}
\Delta X := \sqrt{\sum_{j} P_j(x) [ x_j -x ]^2 } =  \sqrt{ \delta^2X +
    (\bar{x}-x )^2 },
\end{eqnarray} 
where $P_j(x) = \mbox{Tr}[ E_j \rho_x^{\otimes \nu}]$ is the
probability of getting the outcome $j$ when measuring $\rho_x^{\otimes
  \nu}$, $\bar{x} := \sum_{j} P_j(x) x_j$ is the average of the
estimator function, and where
\begin{eqnarray}
\delta^2 X := \sum_{j} P_j(x) [ x_j -  \bar{x}]^2\;,
\end{eqnarray} 
is the variance of the random variable $x_j$. The estimation is said
to be unbiased if $\bar{x}$ coincides with the real value $x$,
i.e.~$\bar{x}=x$, so that, in this case, $\Delta X$ coincides with
$\delta X$.  General estimators however may be biased with $\bar{x}\neq
x$, so that $\Delta X > \delta X$ (in this case, they are called
asymptotically unbiased if $\bar{x}$ converges to $x$ in the limit
$\nu\rightarrow \infty$).

In the main text we restricted our analysis to pure states of the
probe $\rho=|\psi\rangle\langle \psi|$ and focused on projective
measurements associated to unbiased estimation procedures whose RMSE
$\Delta X$ is independent on $x$.
Here we extend the proof to drop the above simplifying assumptions,
considering a generic (non necessarily unbiased) estimation process
which allows one to determine the value of the real parameter $X$
associated with the non necessarily pure input state $\rho$.

Take two values $x$ and $x'$ of $X$ such that their associated RMSE
verifies the following constraints
\begin{eqnarray}
  &&\Delta X(x) \neq 0\;,\label{pos}\\
  &&|x-x'| =
  ( \lambda+1) [\Delta X(x)+\Delta X(x') ] \;,
  \label{dist}
\end{eqnarray}
for some fixed value $\lambda$ greater than 1 (the right hand side of
Eq.~(\ref{dist}) can be replaced by $\lambda [\Delta X(x) +\Delta
X(x')]$ if the estimation is unbiased). In these expressions $\Delta
X(x)$ and $\Delta X(x')$ are the RMSE of the estimation evaluated
through Eq.~(\ref{defdelta}) on the output states $\rho_x^{\otimes
  \nu}$ and $\rho_{x'}^{\otimes \nu}$ respectively (to include the
most general scenario we do allow them to depend explicitly on the
values taken by the parameter $X$).  In the case in which the
estimation is asymptotically unbiased and the quantum Fisher
information $Q(x)$ of the problem takes finite values, the condition
(\ref{pos}) is always guaranteed by the quantum Cram\'{e}r-Rao
bound~\cite{holevo,helstrom,BRAU96,BRAU94} (but notice that our proof
holds also if the quantum Cram\'{e}r-Rao bound does not apply -- in
particular, we do not require the estimation to be asymptotically
unbiased).  The condition~(\ref{dist}) on the other hand is verified
by any estimation procedure which achieves a reasonable level of
accuracy: indeed, if it is not verified, then this implies that the
interval over which $X$ can span is not larger than twice the average
RMSE achievable in the estimation.

Since the fidelity between two quantum states is the minimum of the
classical fidelity of the probability distributions from arbitrary
POVMs~\cite{nc}, we can bound the fidelity between $\rho_x^{\otimes
  \nu}$ and $\rho_{x'}^{\otimes \nu}$ as follows
\begin{eqnarray}
  F :=\Big[ \mbox{Tr} \sqrt{\sqrt{\rho_x^{\otimes \nu}}
    \rho_{x'}^{\otimes \nu} \sqrt{\rho_x^{\otimes \nu}} }\Big]^2
  \leqslant  \Big[
  \sum_j\sqrt{ P_j(x) P_j(x') } \Big]^2\;,\nonumber\\\label{fid}
\end{eqnarray} 
with $P_j(x) = \mbox{Tr} [ E_j \rho_x^{\otimes \nu}]$ and  $P_j(x') = \mbox{Tr} [ E_j \rho_{x'}^{\otimes \nu}]$.
The right-hand-side of this expression can be bound as
\begin{widetext}
\begin{eqnarray}
 \sum_j\sqrt{ P_j(x) P_j(x') }
 &=&  \sum_{j\in I} \sqrt{ P_j(x) P_j(x') }+  \sum_{j\notin I} \sqrt{
   P_j(x) P_j(x') } \nonumber \\
 &\leqslant&   \sqrt{\sum_{j\in I}{ P_j(x) } \sum_{j'\in I}{
     P_{j'}(x') }} +\sqrt{  \sum_{j\notin I} { P_j(x) }
   \sum_{j'\notin I}{P_{j'}(x') } } 
\nonumber \\  &\leqslant& 
  \sqrt{\sum_{j\in I}{ P_j(x) } } +\sqrt{
    \sum_{j'\notin I}{P_{j'}(x') } } \;,\label{boun1}
\end{eqnarray} 
where $I$ is a subset of the domain of possible outcomes $j$ that we
will specify later, and where we used the Cauchy-Schwarz inequality
and the fact that $ \sum_{j'\in I}{ P_{j'}(x') }\leqslant 1$ and $
\sum_{j\notin I} { P_j(x) }\leqslant 1 $ independently from $I$.  Now,
take $I$ to be the domain of the outcomes $j$ such that
\begin{eqnarray}
 |x_j - \bar{x}'| \leqslant \lambda \delta X',
\labell{lam}\;
\end{eqnarray}
where $\lambda$ is a positive parameter (here $\bar{x}'$ and $(\delta X')^2$ are the average and the variance value of $x_j$ computed with the probability distribution $P_{j}(x')$).  
From the Tchebychev inequality it then follows that 
\begin{eqnarray}
 \sum_{j'\notin I}{P_{j'}(x') } \leqslant 1 /\lambda^2\;,\label{asa1}
 \end{eqnarray} 
 which gives a significant bound only when $\lambda>1$.  
 To bound the other term on the rhs of Eq.~(\ref{boun1}) we 
 notice that $|x-x'|  \leqslant \big|x- \bar{x} \big| + \big| x'- \bar{x'}\big| + \big|\bar{x}-\bar{x}'\big|$ and use 
  Eq.~(\ref{dist})  and (\ref{defdelta})  to write 
\begin{eqnarray} 
\big|\bar{x}-\bar{x}' \big|  &\geqslant& (\lambda +1) (\Delta X + \Delta X') -   \big|x- \bar{x} \big| - \big| x'- \bar{x}'\big|  \nonumber \\
&=& (\lambda +1) (\Delta X + \Delta X') - \sqrt{\Delta^2X - \delta^2 X} - \sqrt{\Delta^2X' - \delta^2 X'} \geqslant \lambda (\Delta X + \Delta X')\;.
\end{eqnarray} 
From Eq.~(\ref{lam}) we also notice that  for $j\in  I$  we have 
\begin{eqnarray} 
 \big|\bar{x}-\bar{x}'\big|  
  \leqslant  \big|\bar{x}- x_j\big| + \big|  x_j - \bar{x}'\big| 
  \leqslant  \big|\bar{x}- x_j\big| + \lambda \delta X'\;,
 \end{eqnarray} 
 which  with the previous expression  gives us  
\begin{eqnarray} 
 \big|\bar{x}- x_j\big| \geqslant \lambda (\Delta X + \Delta X') 
 - \lambda \delta X'\geqslant \lambda \Delta X \geqslant \lambda \delta X\;, 
 \end{eqnarray} 
and hence (using again the Tchebychev inequality) 
 \begin{eqnarray}
 \sum_{j\in I}{P_{j}(x) } \leqslant 1 /\lambda^2\;.\label{17}
 \end{eqnarray} 
 Replacing \eqref{asa1} and \eqref{17} into (\ref{fid}) and
 (\ref{boun1}) we obtain
 \begin{eqnarray}\label{fidnew}
F \leqslant 4 /\lambda^2 \;.
\end{eqnarray} 
We can now employ the quantum speed limit inequality~(\ref{newqsl})
from \cite{qspeed} and merge it with the condition (\ref{dist}) to
obtain
 \begin{eqnarray}
   (\lambda +1 )(\Delta X + \Delta X')=  |x'-x|  &\geqslant&
   \frac{\pi}{2} \max \left\{ \frac{\alpha(F)}{\nu(\<H\>- E_0)},
     \frac{\beta(F)}{\sqrt{\nu} \Delta H }\right\} 
   \geqslant \frac{\pi}{2} \max \left\{
     \frac{\alpha(4/\lambda^2)}{\nu(\<H\> - E_0)},
     \frac{\beta(4/\lambda^2)}{\sqrt{\nu} \Delta H}\right\},\label{ila}
\end{eqnarray}
\end{widetext} 
where, as in the main text, we used the fact that $\alpha$ and $\beta$
are decreasing functions of their arguments, and the fact that the
expectation and variances of $H$ over the family $\rho_x$ is
independent of $x$ (since $H$ is independent of $x$).  The first term
of Eq.~\eqref{ila} together with the first part of the $\max$ implies
Eq.~(\ref{bd1}), choosing $\kappa= \sup_{\lambda} \pi
\alpha(4/\lambda^2)/[4(\lambda+1)]\simeq 0.074$, which for unbiased
estimation can be replaced by $\kappa=\sup_{\lambda} \pi
\alpha(4/\lambda^2)/[4\lambda] \simeq 0.091$.  In the case in which
$\Delta X(x) =\Delta X(x')=\Delta X$ we then immediately obtain the
bound~(\ref{ris}).

\begin{figure}[t]
\begin{center}
\epsfxsize=.5\hsize\leavevmode
\epsffile{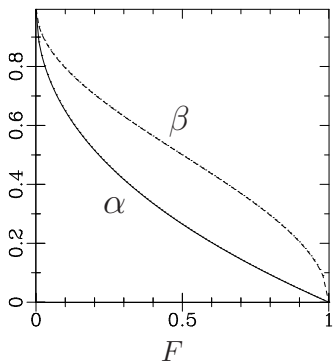}
\end{center}
\caption{ Plot of the functions $\alpha(F)$ and $\beta(F)$ appearing
  in Eq.~(\ref{newqsl}).}
\labell{f:qsl}\end{figure}

\begin{figure}[hbt]
\begin{center}
\epsfxsize=.6\hsize\leavevmode
\epsffile{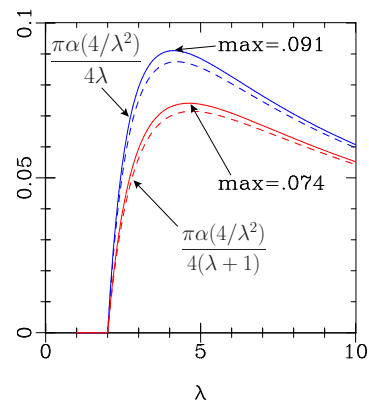}
\end{center}
\caption{Plot of the function $\pi\: \alpha(4/\lambda^2)/(4\lambda)$
  as a function of $\lambda$ (blue continuous line). The function
  $\alpha$ is evaluated numerically according to the prescription of
  \cite{qspeed}. The same function obtained by approximating $\alpha$
  as $\beta^2(F)=4\arccos^2(\sqrt{F})/\pi^2$ is plotted with a blue
  dashed line. The red continuous and dashed lines are analogous, but
  depict the function $\pi\: \alpha(4/\lambda^2)/[4(\lambda+1)]$ that
  must be employed in the case of biased measurements. }
\labell{f:alpha}\end{figure}


\begin{references}
\bibitem{heisenberg}Heisenberg, W., \"Uber den anschaulichen Inhalt der
  quantentheoretischen Kinematik und Mechanik, {\em Zeitschrift f\"ur
    Physik} {\bf 43}, 172-198 (1927), English translation in Wheeler
  J.A.  and Zurek H. eds., {\em Quantum Theory and Measurement}
  (Princeton Univ.  Press, 1983), pg.  62-84.
\bibitem{robertson} Robertson, H.P., The Uncertainty Principle, {\em
    Phys. Rev.} {\bf 34}, 163 (1929).
\bibitem{holevo} Holevo, A.S., Probabilistic and Statistical Aspect
  of Quantum Theory.   (Edizioni della Normale, Pisa 2011).
\bibitem{helstrom} Helstrom, C.W., Quantum Detection and Estimation
  Theory.  (Academic Press, New York, 1976).
    \bibitem{BRAU94} Braunstein, S.L. \& Caves, C.M., Statistical
  distance and the geometry of quantum states.  {\em Phys. Rev. Lett.}
  {\bf 72}, 3439 (1994).
\bibitem{BRAU96} Braunstein, S.L., Caves, M.C. \& Milburn, G.J.,
  Generalized Uncertainty Relations: Theory, Examples, and Lorentz
  Invariance.  {\em Annals of Physics} {\bf 247}, 135-173 (1996).
\bibitem{review} Giovannetti, V., Lloyd, S. \& Maccone, L., Advances
  in Quantum Metrology, {\em Nature Phot.} {\bf 5}, 222 (2011).
\bibitem{caves} Braunstein, S.L., Lane, A.S., \& Caves, C.M.,
  Maximum-likelihood analysis of multiple quantum phase measurements,
  {\em Phys. Rev. Lett.}  {\bf 69}, 2153-2156 (1992).
\bibitem{yurke} Yurke, B., McCall, S.L. \& Klauder, J.R., Phys. Rev.
  A {\bf 33}, 4033 (1986).
\bibitem{barry}Sanders, B.C. \& Milburn, G.J., Optimal Quantum
  Measurements for Phase Estimation, {\em Phys. Rev. Lett.} {\bf 75},
  2944-2947 (1995).
\bibitem{ou} Ou, Z.Y., Fundamental quantum limit in precision phase
  measurement, {\em Phys. Rev. A} {\bf 55}, 2598 (1997); Ou Z.Y.,
  Complementarity and Fundamental Limit in Precision Phase
  Measurement, {\em Phys. Rev. Lett.} {\bf 77}, 2352-2355 (1996).
\bibitem{bollinger} Bollinger, J.J., Itano, W.M., Wineland, D.J. \&
  Heinzen, D.J., Optimal frequency measurements with maximally
  correlated states, {\em Phys. Rev. A} {\bf 54}, R4649 (1996).
\bibitem{smerzi}Hyllus, P., Pezz\'e, L. \& Smerzi, A., Entanglement and
  Sensitivity in Precision Measurements with States of a Fluctuating
  Number of Particles, {\em Phys. Rev. Lett.} {\bf 105}, 120501
  (2010).
\bibitem{rivasluis}Rivas, A. \& Luis, A., Challenging metrological
  limits via coherence with the vacuum, {\em preprint}
  arXiv:1105.6310v1 (2011).
\bibitem{dowling} Anisimov, P.M. et al., 
  Quantum Metrology with Two-Mode Squeezed Vacuum: Parity Detection
  Beats the Heisenberg, {\em Phys. Rev. Lett.} {\bf 104}, 103602
  (2010).
\bibitem{zhang}Zhang, Y.R., et al., Heisenberg Limit of Phase
  Measurements with a Fluctuating Number of Photons, {\em preprint}
  arXiv:1105.2990v2 (2011).
\bibitem{GIOV06} Giovannetti, V., Lloyd, S. \& Maccone, L., Quantum
  metrology.  {\em Phys. Rev. Lett.} {\bf 96}, 010401 (2006).
\bibitem{qspeed} Giovannetti, V., Lloyd, S. \& Maccone, L., Quantum
  limits to dynamical evolution, {\em Phys. Rev. A} {\bf 67}, 052109
  (2003).
\bibitem{margolus} Margolus, N. \& Levitin, L.B., The maximum speed of
  dynamical evolution {\em Physica D} {\bf 120}, 188 (1998).
\bibitem{bhatta} Bhattacharyya K., {\em J. Phys. A} {\bf 16}, 2993
  (1983).
  \bibitem{man} L. Mandelstam and I. G. Tamm, J. Phys. USSR {\bf 9}, 249
  (1945).
\bibitem{kok} Zwierz, M., P\'erez-Delgado, C.A., \& Kok, P., General
  Optimality of the Heisenberg Limit for Quantum Metrology, {\em Phys.
    Rev. Lett.} {\bf 105,} 180402 (2010).
\bibitem{erratum}Zwierz, M.,  P\'erez-Delgado, C.A., \&  Kok, P., Erratum:
  General Optimality of the Heisenberg Limit for Quantum Metrology,
  {\em Phys.  Rev. Lett.} {\bf 107,} 059904(E) (2011).
\bibitem{ssw} Shapiro, J.H., Shepard, S.R. \& Wong, F.C., {\em Ultimate
    quantum limits on phase measurement}, Phys. Rev. Lett. {\bf 62},
  2377-2380 (1989).
\bibitem{HAYA10-1} Hayashi, M., Phase estimation with photon number
  constraint. {\em Progress in Informatics} {\bf 8}, 81-87 (2011);
  arXiv:1011.2546v2 [quant-ph].
\bibitem{nc} Nielsen, M.A. \& Chuang, I.L., Quantum Computation and
  Quantum Information (Cambridge Univ. Press, Cambridge, 2004),
  Eq.~(9.77), pg.~412.
\end{references}
\end{document}